\documentclass[sigconf]{acmart}
\usepackage{booktabs} 

\usepackage[T1]{fontenc}
\usepackage[utf8]{inputenc}
\usepackage{url}
\usepackage{breakurl}
\usepackage{graphicx}
\usepackage{booktabs}
\usepackage{array}
\usepackage{microtype}
\DisableLigatures[f]{encoding = *, family = * }
\usepackage{paralist}
\usepackage{enumerate}
\usepackage{tcolorbox}
\usepackage{refcount}

\newcommand{\rqone}{What is the distribution of (un)happiness among software developers?}
\newcommand{\rqtwo}{What are the experienced causes for unhappiness among software developers while developing software?}

\setcopyright{acmlicensed}

\acmDOI{http://dx.doi.org/10.1145/3084226.3084242}

\acmISBN{978-1-4503-4804-1/17/06}

\acmConference[EASE '17]{21st International Conference on Evaluation and Assessment in Software Engineering}{June 15--16 2017}{Karlskrona, Sweden}
\acmYear{2017}
\copyrightyear{2017}

\acmPrice{15.00}

\begin{document}

\title{On the Unhappiness of Software Developers}

\author{Daniel Graziotin}
\affiliation{%
  \institution{Institute of Software Technology}
  \streetaddress{University of Stuttgart}
  \city{Stuttgart} 
  \state{Germany} 
}
\email{daniel.graziotin@informatik.uni-stuttgart.de}

\author{Fabian Fagerholm}
\affiliation{%
  \institution{Department of Computer Science}
  \streetaddress{University of Helsinki}
  \city{Helsinki} 
  \state{Finland} 
}
\email{fabian.fagerholm@helsinki.fi}

\author{Xiaofeng Wang}
\affiliation{%
  \institution{Faculty of Computer Science}
  \streetaddress{Free University of Bozen-Bolzano}
  \city{Bolzano-Bozen} 
  \state{Italy} 
}
\email{xiaofeng.wang@unibz.it}

\author{Pekka Abrahamsson}
\affiliation{%
  \institution{Department of Computer and Information Science (IDI)}
  \streetaddress{NTNU}
  \city{Trondheim} 
  \state{Norway} 
}
\email{pekkaa@ntnu.no}

\begin{abstract}
The happy-productive worker thesis states that happy workers are more productive. Recent research in software engineering supports the thesis, and the ideal of flourishing happiness among software developers is often expressed among industry practitioners. However, the literature suggests that a cost-effective way to foster happiness and productivity among workers could be to limit unhappiness. Psychological disorders such as job burnout and anxiety could also be reduced by limiting the negative experiences of software developers. Simultaneously, a baseline assessment of (un)happiness and knowledge about how developers experience it are missing. In this paper, we broaden the understanding of unhappiness among software developers in terms of (1) the software developer population distribution of (un)happiness, and (2) the causes of unhappiness while developing software. We conducted a large-scale quantitative and qualitative survey, incorporating a psychometrically validated instrument for measuring (un)happiness, with 2\,220 developers, yielding a rich and balanced sample of 1\,318 complete responses. Our results indicate that software developers are a slightly happy population, but the need for limiting the unhappiness of developers remains. We also identified 219 factors representing causes of unhappiness while developing software. Our results, which are available as open data, can act as guidelines for practitioners in management positions and developers in general for fostering happiness on the job. We suggest considering happiness in future studies of both human and technical aspects in software engineering.

\end{abstract}

%
%

%
%

%
%




\maketitle

\section{Introduction}
The need and importance of managing the individuals forming the software development workforce were identified early in software engineering research~\cite{Humphrey1996}. Management and people related challenges grow as the numbers of software companies and  developers increase with the digitalization of existing businesses and startups founded on software from day one~\cite{Rana:2015er}. A practice that has emerged recently is to promote flourishing happiness among workers in order to enact the happy-productive worker thesis~\cite{Zelenski2008}. Notable Silicon Valley companies and influential startups are well known for their perks to developers~\cite{Marino2008}. Recognizing, managing, and improving the happiness of all stakeholders involved in producing software is essential to software company success~\cite{Denning2012b}.

A novel line of research belonging to behavioral software engineering~\cite{Lenberg:2015bj} is emerging, focusing on the relationship between the happiness of developers and work-related constructs such as performance and productivity~\cite{graziotin2015you,Graziotin2014PEERJ,graziotin2015feelings,Fagerholm2015,Fagerholm2014,Muller2015,Ortu2015a}, and software quality~\cite{Khan2010,Destefanis2016}. The empirical evidence indicates that happy developers perform better than unhappy developers~\cite{Graziotin2014IEEESW}. The studies so far, including those by the present authors, imply that managers and team leaders should attempt to foster developer happiness.

There is the other side of the coin, though. Diener~\cite{Diener:1999cl} and Kahneman~\cite{Kahneman:1999ck} have suggested that objective happiness\footnote{We are using the more colloquial term \emph{happiness} instead of \emph{subjective well-being} throughout the paper as it has historical meaning to research in organizational behavior and psychology~\cite{Fredrickson:2001ul}. Furthermore, as our view of \emph{un}happiness contemplates it as the negative component of happiness, we interchange the two terms when dealing with quantifications of developers' (un)happiness.} can be assessed by the difference between experienced positive affect and experienced negative affect. The happiness equation suggests that maximizing happiness can be achieved by maximizing positive experiences of individuals, minimizing their negative experiences, or both.

Software developers are prone to share horror stories about their working experience on a daily basis~\cite{Graziotin2014IEEESW}. Those in managerial positions should attempt to understand the nature and dynamics of unhappiness in the workplace to create programs for preventing dysfunctional responses among employees~\cite{Vecchio:2000ei}. Psychological disorders such as stress and burnout could be reduced by analyzing the negative affective experiences of developers and turning them positive~\cite{Mantyla2016a}. Furthermore, the voice of practitioners should be heard in software engineering research -- software developers want their unhappiness to be voiced out~\cite{Graziotin2014IEEESW,Ford2015}. For the previously stated reasons, there are calls to understand the benefits of limiting negative experiences on the job~\cite{Fredrickson:2001ul,BenZeev:2010uj,Diener:1999cl}.

The current research on software developers' affective experience lacks a baseline estimation of the distribution of happiness among software developers, as well an understanding of the causes of unhappiness that would be based on a broad sample.

In this paper, we echo the previous calls and aim to \emph{broaden the understanding of unhappiness among software developers}. Based on the existing literature, we set the following research questions (RQs).

\begin{compactenum}[RQ1]
\item \emph{\rqone}
\item \emph{\rqtwo}
\end{compactenum}

To answer the RQs, we conducted a large-scale quantitative and qualitative survey of $2\,220$ software developers in which we asked them to voice out causes of happiness as well as unhappiness. We computed the population estimate of happiness, found 219 causes of unhappiness, and showed that the most prevalent causes of unhappiness are external to developers, suggesting that managers and team leaders have a realistic chance of influencing the happiness of software developers at work. We archived the list of causes as open data \cite{Graziotin:2016ck}.


\section{Background and Related Work}
\label{sec:related:background_and_related_work}
What is happiness, and what does it mean to be happy or unhappy? Intuitively, we could associate this question to the sensing of an individual's affect. We begin by discussing affect, emotions, and moods. 

Russell~\cite{Russell2003} has provided a widely agreed definition of \emph{affect} as ``\textit{a neurophysiological state that is consciously accessible as a simple, non-reflective feeling that is an integral blend of hedonic (pleasure--displeasure) and arousal (sleepy--activated) values}'' (p.\ 147). That is, affect is how we feel at any given point in time, about anything, and this feeling is expressed in how pleasant and activated our state of mind is. We have argued elsewhere~\cite{Graziotin2015SSE} that there are several theories and definitions for emotions and moods. For clarity and brevity, we use Russell's~\cite{Russell2003} theory for the present paper, which considers affect as the atomic unit upon which moods and emotions can be constructed. We consider \emph{moods} as prolonged, unattributed affect, and we consider \emph{emotions} as interrelated events concerning a psychological object, i.e., an episodic process of perception of affect that is clearly bounded in time, in line with several other authors, e.g.,~\cite{Fisher2000, Khan2010}.

From a hedonistic point of view, a blend of affect constitutes an individual's \emph{happiness}~\cite{Haybron2001}. Happiness is a sequence of experiential episodes~\cite{Haybron2001} and being happy (unhappy) corresponds with the frequency of positive (negative) experiences~\cite{Lyubomirsky2005}\footnote{Alternative views of happiness exist, e.g., the Aristotelian eudaimonia considers a person happy because (s)he conducts a satisfactory life full of quality~\cite{Haybron2005}. We have also discussed the role of the centrality of affect in~\cite{graziotin2015you}. Current research in psychology supports the affect balance model as a valid approach to quantify happiness}. Frequent positive (negative) episodes lead to feeling frequent positive (negative) affect, which in turn leads to happiness (unhappiness), represented by a positive (negative) \emph{affect balance}~\cite{Diener2010}. In brief, unhappy individuals are those who experience negative affect more often than positive affect, which is a condition that can be detected by a negative affect balance~\cite{Lyubomirsky2005,Diener2010}.

\subsection{Scale of Positive and Negative Experience}
\label{sssec:related:spane}

Recent studies have found several shortcomings in the Positive and Negative Affect Schedule (PANAS)~\cite{Watson1988b}, the most prominent measurement instrument for assessing happiness, in terms of its affect coverage~\cite{Thompson2007, Li2013,Diener2010} and neglect of cultural differences~\cite{Tsai2006, Li2013}. New scales have been developed that attempt to address PANAS' limitations. Diener et al.~\cite{Diener2010} have presented the Scale of Positive and Negative Experience (SPANE), a short scale that assesses the happiness of participants by asking them to report the frequency of their positive and negative experiences during the last four weeks. SPANE has been reported to be capable of measuring positive and negative affect (and happiness) regardless of the sources, mental activation level, or cultural context, and it captures affect from the entire affective spectrum~\cite{Diener2010, Li2013}. Respondents are asked to report on their affect, expressed with adjectives that individuals recognize as describing emotions or moods, from the past four weeks in order to provide a balance between the sampling adequacy of affect and the accuracy of human memory to recall experiences~\cite{Li2013}, as well as to decrease the ambiguity of people's understanding of the scale itself~\cite{Diener2010}. 

SPANE has been validated to converge to other similar measurement instruments, including PANAS~\cite{Diener2010}. The scale provides good psychometric properties (validity and reliability) which were empirically demonstrated in several large-scale studies~\cite{Diener2010,Silva2013, Li2013, Sumi2013,Jovanovic:2015hw,Corno:2016cy,duPlessis2016}. The scale has been proven consistent across full-time workers and students~\cite{Silva2013}. For these reasons (and for its brevity), we chose SPANE for the purpose of our research.

SPANE is a 12-item scale divided into two sub-scales of positive (SPANE-P) and negative (SPANE-N) experiences. The answers to the 12 items are given on a five-point scale ranging from 1 (\textit{very rarely or never}) to 5 (\textit{very often or always}). The SPANE-P and SPANE-N measures are the sum of the scores given to their respective six items, each ranging from 6 to 30. The two scores are further combined by subtracting SPANE-N from SPANE-P, resulting in the \emph{Affect Balance (SPANE-B)} score. SPANE-B is an indicator of the happiness caused by how often positive and negative affect has been felt by a participant. SPANE-B ranges from $-24$ (\textit{completely negative}) to $+24$ (\textit{completely positive}).

\subsection{Related Studies}
\label{ssec:related:related_studies}

Interest in studying the affect of software developers has risen considerably in the last five years, although we have just started to understand the tip of the iceberg~\cite{Short2015}. To our knowledge, no studies have offered an estimation of the happiness distribution of developers, and only a small number of causes of developers' affect have been examined in a few studies. The present study addresses this research gap.

There are some initial indicators regarding developers' happiness. Generally speaking, studies indicate a positive relationship between the happiness of developers and their performance. Graziotin et al.~\cite{Graziotin2014PEERJ} performed a quasi-experiment on the impact of affect on analytic problem solving and creative performance. The study itself was about consequences of happiness, thus not particularly related to the present article. Yet, Graziotin et al.\ observed that the sample distribution of happiness, measured using SPANE, was significantly greater than 0 (SPANE-B mean=$7.58$, 95\% CI [5.29, 9.85]; median=$9$). The authors noted that the SPANE-B distribution did not resemble a normal distribution. However, the sample was very limited (N=$42$ BSc and MSc students of the same CS faculty); further exploration and validation were suggested based on the observations. We build on these initial observations in the present study.

Some studies have attempted to uncover issues related to affect using both qualitative and quantitative approaches with different degrees of automation. De Choudhury and Counts~\cite{DeChoudhury2013a} investigated the expression of affect through the analysis of $204\,000$ micro-blogging posts from $22\,000$ unique users of a Fortune 500 software corporation. The sentiment analysis revealed that IT-related issues were often sources of frustration. Day-to-day demands, e.g., meetings, were also associated with negative affect.

Ford and Parnin~\cite{Ford2015} have explored frustration in software engineering through practitioner interviews. 67\% of the 45 participants reported that frustration is a severe issue for them. As for the causes for such frustration, the authors categorized the responses as follows: not having a good mental model of the code (for the category ``mapping behavior to cause''), learning curves of programming tools, too large task size, time required for adjusting to new projects, unavailability of resources (e.g., documentation, server availability, \ldots), perceived lack of programming experience, not fulfilling the estimated effort for perceived simple problems, fear of failure, internal hurdles and personal issues, limited time, and issues with peers.

Graziotin et al.~\cite{graziotin2015you} conducted a qualitative study for constructing  an explanatory process theory of the impact of affect on development performance. The theory was constructed by coding data coming from interviews, communications, and observations of two software developers working on the same project for a period of 1.5 months. The theory was built upon the concepts of events, affect, focus, goals, and performance. The study theorized the construct of \emph{attractors}, which are affective experiences that earn importance and priority to a developer’s cognitive system. Attractors were theorized to have the biggest impact on development performance. Finally, the study suggested that interventions (e.g., facilitating reconciliation between developers who are angrily arguing) can mediate the intensity of existing negative affect and reduce their intensity and disruption.

Wrobel~\cite{Wrobel2013} conducted a survey with 49 developers, assessing the participants' emotions that were perceived to be those influencing their productivity. The results showed that positive affective states were perceived to be those enhancing development productivity. Frustration was perceived as the most prevalent negative affect, as well as the one mostly deteriorating productivity.

Ortu et al.~\cite{Ortu2015a}, Destefanis et al.~\cite{Destefanis2016}, and M{\"a}ntyl{\"a} et al.~\cite{Mantyla2016a} conducted a series of mining software repositories studies to understand how affect, emotions, and politeness are related to software quality issues. The studies showed that happiness in terms of frequent positive affect and positive emotions was associated with shorter issue fixing time. Issue priority was found to be associated with the arousal mental activation level, which is often associated with anxiety and burnout.

\section{Method}
\label{sec:method:method}

We employed a mixed research method, comprising both elements of quantitative and qualitative research~\cite{Creswell2009}. In particular, we opted to approach RQ1 with a quantitative investigation, while we addressed RQ2 with a mostly qualitative inquiry. As our aim was to learn from a large number of individuals belonging to a particular population, we considered a survey, implemented as an online questionnaire, to be the most appropriate instrument~\cite{Easterbrook2008}.

\subsection{Sampling Strategy}
\label{sec:method:data-retrieval}

We consider a software developer to be a person concerned with any aspect of the software construction process (such as research, analysis, design, programming, testing, or management activities), for any purpose including work, study, hobby, or passion. Generalizing to the population of software developers is a challenge, because we do not accurately know how many software developers exist in the world and how to reach them. We relied on the GitHub social coding community as a source that fits our purpose of generalization \textit{well enough}, in line with several previous studies (e.g.,~\cite{Gousios:2016hj}). GitHub is the largest social coding community with more than 30 million visitors each month~\cite{Doll:2015uv}, many of which are software developers working on open source and proprietary software ranging from solo work to companies and communities.

To obtain the contact information of GitHub developers, we retrieved related data through the GitHub Archive~\cite{Grigorik:2016}, which stores the collections of \textit{public} events occurring in GitHub. In order to ensure a sample of high quality and variety, we obtained six months of archive data, from March 1 to September 30, 2014. We extracted \textit{unique entries that provided an e-mail address}. We gathered $456\,283$ entries of contact data, which included email address, given name, company, and location of the developers, and the repository name related to the public activity. 41.7\% of the data provided an entry for the \textit{company} field.

\subsection{Survey Design}

We collected data and enhanced the survey in four rounds. During the first three rounds, we piloted the questionnaire design with a limited random sample of contact data (N=$100$ in each round). We discarded the pilots' contact data and questionnaire data from the final survey as many guidelines recommend (e.g.,~\cite{Leon:2011kl}).

The three pilot rounds allowed us to estimate and improve the participation and response rate of the study through a refinement of the questions and invitation e-mail. Through the pilot tests, we understood that we could expect a high percentage of delivered e-mails (between 97\% and 98\%) and that we could expect a low participation rate (between 2\% and 4\%). The participation rate increased in each run.

The questionnaire used in the final survey is composed of (1) questions to collect demographic information, (2) one question carrying SPANE's 12 scale items, and (3) two open-ended questions on the causes of happiness and unhappiness (in terms of SPANE-P and SPANE-N components, see Section \ref{sssec:related:spane}) while developing software. The questionnaire also provides an open-ended field for further comments and a field for optionally leaving an e-mail address for possible follow-ups\footnote{We designed the invitation e-mail and questionnaire text ensuring informed consent by the participants.}. The questionnaire is available in an archived online appendix~\cite{Graziotin:2016ck}.

For estimating the sample size required to make inferences regarding our population of developers, we evaluated the sample size estimations by Bartlett et al.~\cite{Bartlett2001}, Krejcie \& Morgan~\cite{Krejcie:1970io}, Cochran~\cite{Cochran:1977ih}, and Yamane~\cite{Yamane:1967tq}, for a-priori statistical power. None of the authors have proposed sample size estimations for open-ended, qualitative entries. Therefore, we decided to opt for the most conservative settings, i.e., Yamane's simplified formula~\cite{Yamane:1967tq} with $\alpha = .01$ assuming a 2\% response rate. Our calculations resulted in a desirable sample of N=664 complete responses, which we expected to reach with $33\,200$ requests under a 2\% response rate.

We designed and published the questionnaire with \textit{eSurvey Creator} and invited the participants via e-mail. We did not share the survey elsewhere.

\subsection{Analysis and Data Cleaning}

In order to answer RQ1, we needed to describe the distribution of the SPANE-B happiness score (see Section \ref{sssec:related:spane}) and to provide an estimation for the population mean and median. We expected to employ non-parametric methods for the mean and median estimation, given our earlier study~\cite{Graziotin2014PEERJ} and the information obtained from the three pilot runs.

In order to answer RQ2, we developed a coding strategy for the open-ended questions. We applied open coding, axial coding, and selective coding, as described by Corbin and Strauss~\cite{Corbin2008}, as follows. The first three authors individually open coded the same set of 50 random responses using a line-by-line strategy. We met through online video calls in order to compare the coding structure and strategy to reach an agreement, that is, a shared axial coding mechanism. Our unit of observation and analysis, the individual developer, was the starting point. We framed our construction of theoretical categories based on Curtis et al.~\cite{Curtis:1988:FSS:50087.50089} model of constructs that are \emph{internal} or \emph{external}, with the internal group being the developer's own being and the external group having the artifact, process, and people as subcategories. Then, we divided the responses and proceeded to open code them (each coder coded a third of the answers). We held a weekly meeting to follow progress and further discuss the coding structure and strategy. Finally, we merged the codes and performed a final selective coding round. We used \textit{NVIVO~11} for the entire qualitative task. We provide a working example of the various coding phases in the online appendix~\cite{Graziotin:2016ck}.

Data cleaning happened during all stages of the study. We adopted common data cleaning strategies, such as outlier analysis (for example, we examined birth years after 2000 and excluded two 1-year old participants), and excluding participants who were not in the intended population or put random text in the text fields. We list the data inclusion and exclusion criteria in the archived online appendix~\cite{Graziotin:2016ck}. We used R~\cite{R:2016} scripts for supporting and automating the data cleaning, data exploration, and data analysis steps.

\section{Results}
\label{sec:results:results}

This section details the results of our investigation. We first provide descriptive statistics on the sample demographics. Then, we provide the results related to each research question.

\subsection{Descriptive Statistics}
\label{ssec:results:descriptive}
Following our conservative strategy, we randomly sampled $33\,200$ entries from our contact list. Our sending tool delivered $31\,643$ (96.6\%) invitation e-mails; the remaining addresses were either malformed or bounced. $2\,220$ individuals participated (7\% response rate). $1\,908$ participants provided valid data for answering RQ1 (86\%) while $1\,318$ provided valid data to provide answers to RQ2 (59\%). Based on the pilots, we anticipated that some participants would leave the open-ended questions unanswered. Our sampling strategy paid off: we exceeded the required threshold (N=664) for generalizing. The rich sample offered us the opportunity to stay conservative for analyzing the data, too. We could minimize bias by retaining only the data provided by the participants that completed the entire questionnaire. That is, we kept N=1318 for answering all our RQs.

Our sample of N=1318 responses resulted in $1\,234$ male participants (94\%) and $65$ female (5\%). The remaining $19$ participants declared their gender as \textit{other / prefer not to disclose}. The average year of birth was 1984 (standard deviation (sd)=9.33), while the mean was 1986. There was diversity in terms of nationality, with 88 countries. The most represented nationalities were American (24\%), Indian (6\%), Brazilian (6\%), Russian (5\%), and British (4\%).

A total of 993 (75\%) of the participants were professional software developers, 15\% of the sample were students, and 8\% were in other roles (such as manager, CEO, CTO, and academic researcher). The remaining participants were non-employed and not students.

The participants declared an average of 8.29 years (sd=7.77) of experience with working with software development, with a median of 5 years. 240 participants developed software either as a hobby, passion, or volunteer without pay, 161 participants worked either as freelancer or consultant in companies, 105 participants were a one-person company or self-employed in a startup, and 812 were employed in a company or a public organization. The reported size of the participants' company or organization also varied considerably, with 13.3\% of the participants working alone, 33.6\% in small entities (2-10 persons), 34.4\% in medium entities (11-250), and 18.7\% in large to very large entities (250-5000 and more).




Regarding the qualitative data, we reached a total of 590 codes in the initial coding phases. After the merge and cleanup phases, we obtained 219 codes that were referenced $2\,280$ times in text (average of 10.41 references per code).

\subsection{RQ1---What is the Distribution of (Un) \\ Happiness Among Software Developers?}
\label{ssec:resultsrq1}
Our sample of N=1318 participants had a SPANE-B (see Section \ref{sssec:related:spane}) mean score of 9.05 (sd=6.76), a median score of 10, and a range of [-16, 24].

\begin{figure}
\centering
\includegraphics[width=\columnwidth]{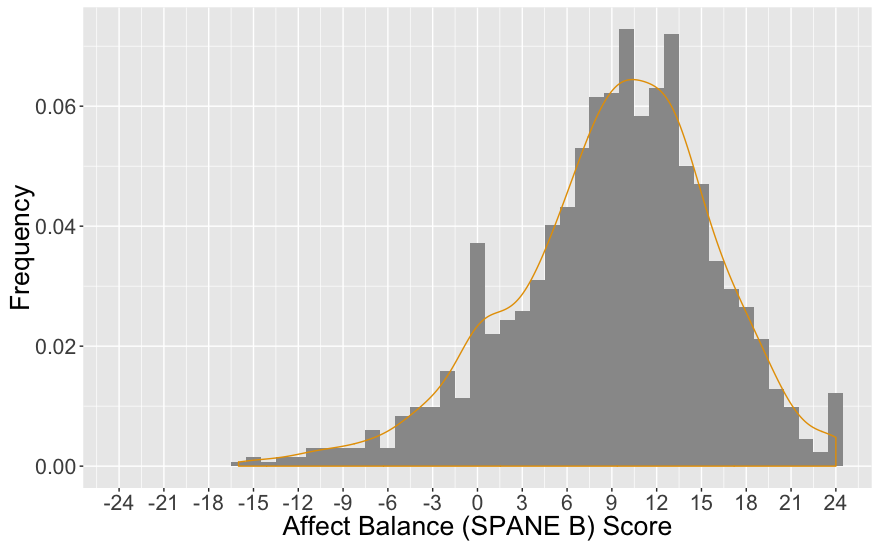}
\caption{Happiness of software developers (SPANE-B distribution).}
\label{fig:results:happiness_of_devs}
\end{figure}

We followed the recent suggestion by Kitchenham et al.~\cite{Kitchenham:2016bm} to use kernel density plot instead of boxplots. The plot of the SPANE-B score is shown in Figure \ref{fig:results:happiness_of_devs}. The plot indicates a likely non-normal distribution of the data, as expected. A description of the SPANE-B score by the \textit{psych} R package~\cite{Psych:2016} showed a skew of -0.53 and a kurtosis of 0.46, indicating a slightly asymmetrical distribution with a long tail to the left that is flatter than a standard normal distribution. Strong evidence for non-normality of the data was supported by a Shapiro-Wilk test for normality ($W=0.98, p<0.0001$).

We estimated the population's true mean for SPANE-B via bootstrapping as $9.05$ ($2000$ replications, 95\% CI [8.69, 9.43]). We estimated the population's true median for SPANE-B via bootstrapping as $10$ ($2000$ replications, 95\% CI [9.51, 10.71]). We show in the online appendix~\cite{Graziotin:2016ck} that estimating those values with the expanded sample of N=1908 would yield similar results.

\subsection{RQ2---What are the Experienced Causes for Unhappiness Among Software Developers While Developing Software?}
\label{ssec:resultsrq2}

We plotted the demographic data gathered from the questionnaire, and compared it with the SPANE-B value. None of the quantitative data plots indicated a relationship with the happiness of developers. This includes variables such as gender, age, nationality, working status, company size, percentage of working time dedicated to developing software, and monthly income. Thus, we conclude that they are not the primary determinants of unhappiness. This further confirmed our original research design to use qualitative data to explore the causes. We identified 219 causes of unhappiness, which are grouped into 18 categories and sub-categories (including the top category of causes of unhappiness while developing software). We report here only the main emerged categories and top 10 factors. Because of space limitations, we provide the demographic data plots and our complete coding as open data in the online appendix~\cite{Graziotin:2016ck}. 

\subsubsection{Main Categories}
\label{sssec:resultsrq2:maincategories}
The main types of factors causing unhappiness among software developers are organized under two main categories. The causes of unhappiness internal to individual developers, directly related to their personal states, or originated by their own behaviors, are classified under the \textit{developer's own being} category. These occurred a total of $437$ times. In contrast, \textit{external causes} are the causes of unhappiness external to individual developers, by which developers are affected but have little or no control of. The total occurrence of external causes is $1\,843$ times. This indicates that developers are much more prone to experiencing and recalling externally-provoked unhappy feelings than internally generated ones.

The \textbf{developer's own being} (i.e., internal causes) category contains 22 internal factors. These factors do not demonstrate a clear structure. This to some extent reflects the versatile states of mind of developers and the feelings they could have while they develop software.

The factors in the \textbf{external causes} category are further divided into the sub-categories shown in Table \ref{tbl:results:categories_external}.
\textit{People-related factors}: the external causes of unhappiness related or attributable to people whom developers interact with, to their characteristics or behaviors. These occurred 416 times and are further divided based on the roles of the people.
\textit{Artifact and working with artifact}: the external causes of unhappiness related to artifacts in software development projects and developers' interactions with them occurred 788 times. The causes are further grouped based on the types of artifacts that developers are dealing with. 
\textit{Process-related factors}: the external causes of unhappiness related to issues in the management of software development process and day-to-day work. This type of causes occurred 544 times.
\textit{Other causes}: the external causes of unhappiness not classified under any of the above-mentioned ``external factors'' categories. These non-specific causes occurred 95 times. 

\begin{table}[t]
\centering
\caption{Categories for External Causes of Unhappiness}
\label{tbl:results:categories_external}
\begin{tabular}{ll}
\toprule
\textbf{Main category} & \textbf{Sub-categories} \\
\midrule
People (416) & \begin{tabular}[c]{@{}l@{}}Colleague (206)\\ Manager (122)\\ Customer (49)\end{tabular} \\
\midrule
\begin{tabular}[c]{@{}l@{}}Artifact and \\ working with artifact\\ (788)\end{tabular} & \begin{tabular}[c]{@{}l@{}}Code and coding (217)\\ Bug and bug fixing (194)\\ Technical infrastructure (151)\\ Requirements (99)\end{tabular} \\
\midrule
\begin{tabular}[c]{@{}l@{}}Process-related \\ factors (544)\end{tabular} & No sub-categories  \\
\midrule
Other causes (95) & No sub-categories \\
\bottomrule
\end{tabular}
\end{table}

\subsubsection{10 Most Significant Causes of Unhappiness}
\label{sssec:resultsrq2:topfactors}
We extracted a list of 10 factors that occurred most often in the survey responses as the causes of unhappiness. They are listed in Table \ref{tbl:results:top10_unhappiness}.

\begin{table*}
\caption{Top 10 Causes of Unhappiness, Categories, and Frequency}
\label{tbl:results:top10_unhappiness}
\centering
\begin{small}
\begin{tabular}{llr}
\toprule
Cause & Category &  Freq. \\
\midrule
Being stuck in problem solving & software developer's own being & 186 \\
Time pressure & external causes$\,\to\,$process & 152 \\
Bad code quality and coding practice  & external causes$\,\to\,$ artifact and working with artifact$\,\to\,$ code and coding  & 107 \\
Under-performing colleague  & external causes$\,\to\,$ people$\,\to\,$ colleague  & 71 \\
Feel inadequate with work  & software developer's own being & 63 \\
Mundane or repetitive task  & external causes$\,\to\,$ process & 60 \\
Unexplained broken code & external causes$\,\to\,$ artifact and working with artifact$\,\to\,$ code and coding & 57 \\
Bad decision making & external causes$\,\to\,$ process & 42 \\
Imposed limitation on development  & external causes$\,\to\,$ artifact and working with artifact$\,\to\,$ technical infrastructure  & 40 \\
Personal issues -- not work related & software developer's own being  & 39 \\
\bottomrule
\end{tabular}
\end{small}
\end{table*}

Three of these top 10 causes are part of \textit{software developer's own being}. \textbf{Being stuck in problem solving} is by far the most significant among the three factors. Software development is essentially composed of problem-solving activities, often intellectually demanding. It is common that developers may be stuck in coding, debugging and all sorts of other tasks. As one respondent commented: ``\textit{I feel negative when I get really stuck on something and cannot get around it}''. Another respondent elaborated: ``\textit{I also thought of situations where I'm debugging some issue with the code and I can't figure out why it isn't working -- when it seems like everything should work, but it just doesn't. This is definitely one of the biggest gumption traps I encounter}''.

Another significant internal cause is a \textbf{feeling of inadequate skills or knowledge}, as shown in this response: ``\textit{Once I encountered hashmap, and I couldn't understand it while I knew it is important. I felt frustrated and afraid}''. The inadequate feeling can be manifested as feeling unskilled in certain aspects of the work, feeling under-qualified with respect to the task given, or feeling a lack of familiarity with tools, languages, frameworks, or development methods that are used in the projects. 

The third significant cause related to developer's own being is not related to work, but \textbf{personal issues}. Software developers are not living in vacuum while working on their software projects, and often non-work related, personal or private issues may affect them and cause their unhappy feelings during work. ``\textit{I never feel 100\% productive when there's something from my private life bugging me. No, I'm not a robot, I'm human, and can't forget the rest of the world when I start IntelliJ up}''. Among the non-work related issues, family related issues are most frequently mentioned: ``\textit{Family related issues has huge impact on my feeling while working, I feel down and can't achieve the goals I set for my work day}''. 

The seven remaining most significant factors are all \textit{external}. Among \textit{people-related} causes, the \textbf{under-performance of colleagues}, either team members, colleagues in other teams, or external collaborators, most often make developers experience negative feelings and affect their work consequently. An illustrative episode is reported in this response: ``\textit{Last time I felt angry when a senior developer again committed an update ruining a beautiful generic solution I've made before. It was easy to refactor it, but his ignorance or routine annoyed me}''. Software development is often teamwork. It is frequently frustrating to a developer to see that other colleagues do not spend time to keep up to speed with modern development technology and practices.

It comes as no surprise that \textbf{bad code quality and coding practices} make developers unhappy. In almost all cases, bad code was a cause of unhappiness if it was written by other developers: ``\textit{Sad/angry when reading others' code that I have to use and I realize it is full of bugs}''; ``\textit{having encountered a particularly bad (unreadable, poorly formatted, not commented at all, badly structured) piece of code written by another developer that I had to work on}''. Only a few participants reported unhappiness caused by ``\textit{poorly written code (often by past-me)}''. That is, unhappiness from code written by the participants themselves was raised only when regretting past code.

Another significant factor related to code that makes developers feel bad is when they could not explain why the code is not working as it is supposed to (\textbf{unexplained broken code}): ``\textit{When you haven't changed the code, and suddenly the project doesn't compile anymore. Worst feeling ever. (Afraid/sad/angry)}''.

Apart from code, issues in the technical infrastructure a software project relies on often contribute to negative feelings among developers, especially when it is supposed to support software development, but instead imposes constraints or limitations (\textbf{imposed limitation on development}). One respondent described this situation perfectly: ``\textit{Angry happens quite often because tools, programming languages, etc. don't do as expected. Sometimes because they are buggy, sometimes there are some limitations in them (by design/by ignoring or not considering enough use cases, etc.), which makes one need to find work-arounds/mess-up otherwise clean code, repeat code unnecessarily etc}''. 

Regarding the top significant causes related to general software development \textit{process}, the respondents consider that high \textbf{time pressure} they feel, often generated by ``\textit{unrealistic}'', ``\textit{unjustified}'' and ``\textit{crazy}'' deadlines, will almost surely push them into very unhappy states. A respondent described vividly this situation: ``\textit{I remembered a day. I have a lot of phone call from my boss to done a project. in that situation time was running and project move slowly and phone every a minute ringed}''.

Contrasting the image of high time pressure, with hectic rushing towards deadlines, is working on \textbf{mundane or repetitive tasks}, which is another process-related factor that often causes negative feelings of developers. ``\textit{Tedious}'', ``\textit{boring}'', ``\textit{dull}'', ``\textit{monotonous}'', ``\textit{trivial}'', ``\textit{recurrent}'', etc., are the words the respondents used to describe the tasks that make them unhappy. ``\textit{I tend to feel negative or bad when I am doing something that is not challenging, I instantly feel sleepy and bored}'', a respondent stated.

\textbf{Bad decision making} is yet another process-related factor that often leaves developers in an unhappy state. More than bad business decisions, developers are more often affected (emotionally as well) by bad technical decisions made by their superiors or peers. One example depicts such a scenario: ``\textit{Generally negative emotions stem from board meetings where an executive or coworker makes an uninformed or ill-advised decision that causes a `dirty' codebase change}''. Bad decision making is also perceived by developers if they are not involved in decision making processes.

\section{Discussion}\label{sec:discussion:discussion}

Through our analysis to answer RQ1 (Section~\ref{ssec:resultsrq1}), we estimated the population mean SPANE-B score to be $9.05$, indicating a happiness balance on the positive side (see Section~\ref{sssec:related:spane}). In terms of the norms reported in Diener et al.~\cite{Diener2010}, this result is in the 65th percentile, indicating that the happiness balance is also higher than what could be expected in a larger human population. The various psychometric studies of SPANE report sample means but no confidence intervals, meaning that the best comparison possible is through means and standard deviations. Those studies have found mean SPANE-B scores above zero, but several score points lower than in our sample: $7.51$ (sd=$8.21$) in a sample of men and $4.53$ (sd=$8.17$) in a sample of women in Italy~\cite{Corno:2016cy}; $6.69$ (sd=$6.88$) in a sample of college students from five universities and colleges in USA and one university in Singapore~\cite{Diener2010}; $6.66$ (sd=$8.18$) in large sample of more than $21\,000$ employees in the Chinese power industry~\cite{Li2013}; $5.96$ (sd=$6.72$) in a multicultural student sample at a large South African university~\cite{duPlessis2016}; $4.41$ (sd=$7.79$) in a sample of full-time employees and $5.10$ (sd=$7.54$) in a sample of university students, both in Portugal~\cite{Silva2013}; and $4.30$ (sd=$7.50$) in a sample of Japanese college students~\cite{Sumi2013}.

Our findings about the higher-than-expected SPANE-B score confirm and reinforce our previous observations~\cite{Graziotin2014PEERJ} that 
\begin{inparaenum}[(1)]
\item software developers are a slightly happy population, and that 
\item there is no evidence that the distribution of SPANE-B scores for the population of software developers should cover the full range of $[-24, +24]$. 
\end{inparaenum}
This does not mean that software developers are happy to the point that there is no need to intervene on their unhappiness. On the contrary, we have shown that unhappiness is present, caused by various factors and some of them could easily be prevented. Our observations and other studies show that unhappiness has a negative effect both for developers personally and on development outcomes. Furthermore, these results have implications for research, as outlined below.

For answering RQ2, we have shown a wide diversity and weight of factors that cause unhappiness among developers (Section \ref{ssec:resultsrq2}). The causes of unhappiness that are external to developers, and thus more controllable by managers and team leaders, could have an incident rate that is 4 times the one of the factors belonging to the developer's own being. We expected that the majority of the causes of unhappiness would come from human related considerations (416 references); however, technical factors from the artifact (788) and the process (544) dominate the unhappiness of developers, highlighting the importance of strategic architecture and workforce coordination.

Being stuck in problem solving and time pressure are the two most frequent causes of unhappiness, which corroborates the importance of recent research that attempts to understand them~\cite{graziotin2015you,Muller2015,Mantyla2016a}. Lack of experience could explain the prevalence of the first category in some cases, but since software development is inherently about problem solving, and realistic projects include an element of problem solving and learning, it does not seem adequate to explain this result by lack of experience alone. It may be necessary to accept that software development comes with its share of difficult tasks that cannot be avoided. Psychological \textit{grit} could be an important characteristic to train among software developers. Strategies for both coping with the negative feeling associated with being stuck and systematic strategies for actually solving problems in general and specific scenarios can be called for.

Several top causes are related to the perception of inadequacy of the self and others, which encourages recent research activities on intervening on the affect of developers~\cite{graziotin2015you}. Finally, we see that factors related to information needs in terms of software quality and software construction are strong contributors to unhappiness among developers. This reinforces recent research activities on those aspects (e.g., ~\cite{Fritz:2011en}) and encourages proposed~\cite{graziotin2016software} research activities that attempt to merge affective reactions and information needs.

\subsection{Limitations}
We designed our survey with the stated aim of gaining understanding of the characterization of the unhappiness of developers and the causes of unhappiness in software development. We phrased the questions in our survey by following guidelines from the literature~\cite{Creswell2009,Oppenheim1992} and from our prior experience with the research topic~\cite{graziotin2015affect,Graziotin2014IEEESW,Graziotin2014PEERJ,Graziotin2015SSE,graziotin2015you,graziotin2015feelings,Fagerholm2014, Fagerholm2015, Fagerholm2015a}. We phrased the questions to avoid priming specific answers to the respondents. The validation of the questions was through (1) adopting a psychometrically validated measurement instrument for happiness~\cite{Diener2010}, (2) limiting the remaining quantitative questions to a demographic nature, and (3) conducting three pilot runs. We discuss specific threats to validity below.

\subsubsection{Internal Validity--Credibility} 

With respect to the happiness measurement, as reported in Section \ref{sssec:related:spane}, several large scale studies have found good psychometric properties (reliability and validity) for SPANE~\cite{Diener2010,Silva2013, Li2013, Jovanovic:2015hw,Corno:2016cy}, and the instrument was empirically shown to be consistent across full-time workers and students~\cite{Silva2013} and memory recall of events~\cite{Diener2010, Li2013}.

In order to classify the causes of unhappiness of developers, we used a qualitative coding process. Whether causality can be inferred only by controlled experiments is a much-debated issue~\cite{Creswell2009,Djamba:2002jo,Glaser:2013ha}. Several authors, e.g.,~\cite{Glaser:2013ha}, maintain that human-oriented research allows causality to be inferred from the experience of the participants through qualitative data analysis, provided that there is a strong methodology for data gathering and analysis. In this case, our aim was to uncover causes of unhappiness as experienced by developers themselves. Since we extracted the causes from first-hand reports, they should accurately represent the respondents' views. As far as possible, we have remained faithful to these views when categorizing the material. The chain of evidence from source material to results is fully documented and traceable (see the online appendix~\cite{Graziotin:2016ck}). The ratio between reported internal and external causes may be affected by the respondents' ability to correctly attribute their unhappiness. We note that we claim no general relationship between any specific causes and unhappiness; only experienced causes of unhappiness are claimed.

A question-order effect~\cite{Sigelman:1981fh} could have influenced the responses by setting their context. As the present study was conducted in the context of a larger study on both happiness and unhappiness, we randomized the order of appearance of the questions related to affectiveness, thus limiting a potential order effect.

Social desirability bias~\cite{Furnham:1986ec} may have influenced the answers in the sense that participants would attempt to appear in a positive light before the eyes of the enquirer. We limited the bias by informing the participants that the responses would be anonymous and evaluated in a statistical form, and addressing the ethical concerns of the study. In our view, the responses appear candid, indicating that participants have felt comfortable expressing their true views.

\subsubsection{Generalizability--Transferability} 
Our dataset of software developers using GitHub is limited with respect to representativeness of the average software developer. The data set used in this study (see Section \ref{sec:method:data-retrieval}) contains only accounts with public activity during a six-month period. However, it is likely that a significant portion of the inactive accounts are not of interest to this study, as we sought active developers. 

The degree to which our conclusions are generalizable beyond the GitHub population may be limited. For instance, it is possible that the GitHub population is slightly younger than developers in general, and age may explain differences in the degree and nature of unhappiness. Also, the sample may be biased towards people that are more comfortable with displaying their personal performance in public, or face no other kinds of barriers to doing so (e.g.,\ company policy). However, GitHub is a reliable source for obtaining research data, allowing replication of this study on the same or different populations. The GitHub community is large and diverse, with a claim of more than 30 million visitors each month~\cite{Doll:2015uv}, many developing open source and proprietary software, and ranging from solo work to companies and communities. Furthermore, as shown in Section \ref{ssec:results:descriptive}, our sample is well balanced in terms of demographic characteristics, including participant role, age, experience, work type, company size, and student versus worker. By comparing confidence intervals, we did not observe significant differences in terms of the SPANE-B score when varying role (worker, student) or age. This further highlights the validity and reliability of the SPANE measurement instrument and the stability of our dataset. 

Our sample is not evenly balanced in terms of gender, with males being in the vast majority. We believe, however, that our sample is representative to some extent in terms of gender as well, since males are overrepresented in software engineering jobs, likely due to gender bias~\cite{Ortu:2016gz,Terrell:2016dq,Ford:2016}. However, our sample may be extreme in this respect; while exact data is difficult to obtain, some non-academic surveys have shown, e.g., 7.6\%\footnote{StackOverflow Developer Survey 2017, \url{https://stackoverflow.com/insights/survey/2017}}, 16\%\footnote{LinkedIn Blog post ``Women in Software Engineering: The Sobering Stats'', reporting statistics based on LinkedIn data, \url{https://business.linkedin.com/talent-solutions/blog/2014/03/women-in-engineering-the-sobering-stats}}, and 20\%\footnote{Bureau of Labor Statistics Current Population Survey: Annual averages, Software developers, applications and systems software, \url{https://www.bls.gov/cps/cpsaat11.htm}} females, but the numbers can depend on the definition of developer and the countries or cultures represented in the data. A possible explanation is that males are particularly overrepresented among GitHub developers, but more demographic data would be needed to ascertain this. In summary, we consider our sample to be large and diverse enough to warrant claims regarding software developers to the extent possible in a single study. Further replication is necessary to validate the findings and obtain details on demographic subgroups.

\subsection{Recommendations for Practitioners}

Our study has found a plethora of causes of unhappiness of developers that are of interest to practitioners regardless of their roles. We summarized the most prominent ones in the present paper, but practitioners could be interested in the complete list of factors and occurrences that is freely available online as open data~\cite{Graziotin:2016ck}. 

Team members may be interested in the causes of unhappiness for enabling self-regulation and emotional capability mechanisms~\cite{Akgun2011b} for reducing personal and group unhappiness. Knowing what might cause unhappiness in the short and long term could encourage developers to be more considerate towards their peers. For example, it might be worth thinking twice about leaving others to clean up badly written code. For similar reasons, managers should carefully attempt to understand the unhappiness of developers using the present paper as support. Those in leadership positions should attempt to foster happiness by limiting unhappiness. Previous research (e.g.,~\cite{Graziotin2014PEERJ,graziotin2015feelings,Destefanis2016}) has shown that the benefits of fostering happiness among developers are substantial especially in terms of software development productivity and software quality. In a related paper on the consequences of unhappiness of developers, we found that addressing unhappiness could limit damage on different aspects of software development, including developers, artifacts, and development processes~\cite{Graziotin2017}. We believe that the results of the present study will potentially enhance the working conditions of software developers. This is corroborated by previous research~\cite{graziotin2015you} suggesting that intervening on the affect of developers may yield large benefits at low cost. We note that such interventions should consider issues of privacy and cultural differences. Whether to intervene in issues outside the work context is an open question, with possible legal constraints.

Furthermore, the vast majority of the causes of unhappiness are of external type. Since external causes may be easier to influence than internal causes, and since influencing them will impact several developers rather than only one at a time, this suggests that there are plenty of opportunities to improve conditions for developers in practice.

\subsection{Implications for Researchers}

We believe that the results of the present work can spawn several important future research directions. A limited set of the found causes of unhappiness has been investigated previously in the software engineering literature. However, while the previous work offers valuable results, it appears limited either because of being framed too generally in psychology research -- resulting in findings regarding general job performance settings -- or due to a narrow focus on a single emotion (e.g., frustration). The framing offered by the present study sheds new light on these previous studies by considering them in terms of happiness and affect. Here, we suggest three implications for research that we believe are of high importance and priority.

Our result regarding the distribution of happiness among developers suggests that happiness -- in terms of the SPANE instrument score -- is centered around $9.05$, higher than what may be expected based on other studies using the instrument. Our question for future research is to understand whether
\begin{inparaenum}[a)]
\item a higher relativity should be embraced when analyzing the affect of developers and its impact on software engineering outcomes, or
\item developers require tailored measurement instruments for their happiness as if they are a special population.
\end{inparaenum}
Validating the score through replication, and, if it is found to be stable, investigating the reasons for it being higher than in several other populations, are important aims for future research.

As reported in the previous section, most causes of unhappiness are of external type and they may be easier to influence than internal causes. We see that much research is needed in order to understand the external causes and how to limit them. Further understanding of the underlying reasons for the ratio between external and internal causes is also needed.

Finally, the present study highlights how studies of human aspects in software engineering are important for the empirical understanding of how software development can be improved. Many questions in software engineering research require approaches from behavioral and social sciences; we perceive a need in academic discourse to reflect on how software engineering research can be characterized and conducted in terms of such paradigms.

Software engineering studies on human factors often call for further human aspects studies. Yet, we believe that the present study calls for much technical research as well, because the highest source of unhappiness among software developers is related to artifacts and working with artifacts. One example is related to debugging and bug fixing, as they appear often in the causes of unhappiness. This suggests that much research is needed for supporting humans in the maintenance of software, e.g.,\ in terms of information needs and mechanisms for strategic coordination of the workforce and the software architecture. Furthermore, emotional support for the sometimes frustrating and tedious work with software maintenance might increase the quality of results.

\section{Conclusion}

In this paper, we presented a mixed method large-scale survey ($1\,318$ complete and valid responses) to broaden the understanding of unhappiness among software developers. Our key contributions are as follows, and are publicly archived as open access and open data~\cite{Graziotin:2016ck}:

\begin{enumerate}[(C1)]
\item An estimate of the distribution of (un)happiness among software developers.
\item An analysis of the experienced causes for unhappiness among software developers while developing software.
\end{enumerate}

Our results show that software developers are a slightly happy population. The consequences of that result need to be explored in future studies. Nevertheless, the results do not remove the need for limiting the unhappiness of developers, who have repeatedly asked to be given a voice through research and in the design of studies.

The results of our study have also highlighted 219 fascinating factors about the causes of unhappiness while developing software. These should be further explored in future research and used as guidelines by practitioners in management positions and developers in general for fostering happiness on the job. We also call for replications of the study.

\section*{Acknowledgment}
The authors would like to thank all those who participated in this study. Daniel Graziotin has been supported by the Alexander von Humboldt (AvH) Foundation.
%
\bibliographystyle{ACM-Reference-Format}
\small{
       \bibliography{references} 
}
%
%
\end{document}